\documentclass[12pt]{iopart}
\usepackage{graphicx,color,amssymb}
\usepackage{color}
\definecolor{MPI}{rgb}{1,0,0}

\definecolor{GL}{rgb}{0,0,1}

\definecolor{Curro}{rgb}{0,0.5,0}

\definecolor{orange}{rgb}{0.9,0.6,0.1}

\newcommand{\LaOFeAs}       {${\rm La} {\rm O} {\rm Fe} {\rm As}$}
\newcommand{\LaOFFeAs}       {${\rm La} {\rm O}_{0.9} {\rm F}_{0.1} {\rm Fe} {\rm As}$}
\newcommand{\LaOFxFeAs}       {${\rm La} {\rm O}_{1-x} {\rm F}_{x} {\rm Fe} {\rm As}$}
\newcommand{\slrrt}     {$(T_1T)^{-1}$}
\newcommand{\slrrtext}   {spin lattice relaxation rate}
\newcommand{\slrr}      {$T_1^{-1}$}
\newcommand{\f}        {$^{19}$F}

\newcommand{\fe}        {$^{57}$Fe}
\newcommand{\as}        {$^{75}$As}
\newcommand{\la}        {$^{139}$La}

\newcommand{\Kab}      {$K_{ab}$}

\newcommand{\vekg}[1]{\mbox{{\boldmath $#1$}}}

\begin{document}

\title[Electronic properties of LaO$_{1-x}$F$_x$FeAs in the normal state probed by NMR/NQR]{Electronic properties of LaO$_{1-x}$F$_x$FeAs in the normal state probed by NMR/NQR}

\author{H.-J. Grafe$^1$, G. Lang$^1$, F. Hammerath$^1$, D. Paar$^{1,2}$, K. Manthey$^1$, K. Koch$^3$, H. Rosner$^3$, N. J. Curro$^4$, G. Behr$^1$, J. Werner$^1$, N. Leps$^1$, R. Klingeler$^1$, H.-H. Klauss$^5$, F. J. Litterst$^6$, B. B\"uchner$^1$}

\address{$^1$IFW Dresden, Institute for Solid State Research, P.O. Box 270116, D-01171 Dresden, Germany\\
         $^2$Department of Physics, Faculty of Science, University of Zagreb, P. O. Box 331, HR-10002 Zagreb, Croatia\\
         $^3$Max Planck Institute for Chemical Physics of Solids, N\"othnitzer Strasse 40, D-01187 Dresden, Germany\\
         $^4$Department of Physics, University of California, Davis, CA 95616, USA\\
         $^5$Institut f\"ur Festk\"orperphysik, TU Dresden, D-01069 Dresden, Germany\\
         $^6$Institut f\"ur Physik der Kondensierten Materie, TU Braunschweig, D-38106 Braunschweig, Germany}
\ead{h.grafe@ifw-dresden.de}

\begin{abstract}
We report \la , \fe , and \as\ Nuclear Magnetic Resonance (NMR) and Nuclear Quadrupole Resonance (NQR) measurements on powders of the new \LaOFxFeAs\ superconductor  for $x$=0 and $x$=0.1 at temperatures up to 480 K, and compare our measured NQR spectra with LDA calculations.
For all three nuclei in the $x$=0.1 material, it is found that the local Knight shift increases monotonically with increasing temperature, and scales with the macroscopic susceptibility, suggesting a single magnetic degree of freedom.
Surprisingly, the spin lattice relaxation rate for all nuclei also scale with one another, despite the fact that the form factors for each site sample different regions of $\vekg{q}$-space.  This result suggests a lack of any $\vekg{q}$-space structure in the dynamical spin susceptibility that might be expected in the presence of antiferromagnetic correlations.  Rather, our results are more compatible with simple quasi-particle scattering. Furthermore, we find that the increase in the electric field gradient at the As cannot be accounted for by LDA calculations, suggesting that structural changes, in particular the position of the As in the unit cell, dominate the NQR response.
\end{abstract}

%Uncomment for PACS numbers title message
%\pacs{00.00, 20.00, 42.10}
% Keywords required only for MST, PB, PMB, PM, JOA, JOB?
%\vspace{2pc}
%\noindent{\it Keywords}: Article preparation, IOP journals
% Uncomment for Submitted to journal title message
%\submitto{\JPA}
% Comment out if separate title page not required

\maketitle

\section{Introduction}

The recent discovery \cite{Kamihara2008} of superconductivity  in the layered ferropnictides ${\rm R} {\rm O}_{1-x} {\rm F}_{x} {\rm Fe} {\rm As}$ (R = rare earth) has raised great interest within the solid state community.
Not only does the transition temperature, $T_c$, reach a maximum at 55 K, but it is strongly dependent on the rare-earth ion and the pressure \cite{Takahashi2008,Yang2008,Ren2008}.  Furthermore, the normal state properties exhibit some unusual features, which are reminiscent of the copper oxide high temperature superconductors (HTSC). In particular, there is a pseudogap-like decrease of the magnetic susceptibility at low temperatures \cite{Ahilan2008,Grafe2008,Nakai2008,Klingeler2008}, and 3D antiferromagnetic order in the parent (undoped) compound \LaOFeAs\ with $T_N~\sim $ 140 K. An important difference, however, is that the ferropnictides exhibit metallic properties, and clearly are not Mott insulators \cite{Cruz2008,klauss}.
Upon doping, the antiferromagnetic order is destroyed in both families, and a superconducting ground state emerges in the phase diagram.
The proximity of superconductivity to an antiferromagnetic ground state and the appearance of the pseudogap features hint at the presence of magnetic correlations in these materials, as in high $T_c$ cuprates, that may play a critical role in the underlying physics of the superconductivity.

In this article, we report Nuclear Magnetic Resonance (NMR), and Nuclear Quadrupole Resonance (NQR) investigations of the normal state of \LaOFFeAs\ and \LaOFeAs.
Our results shed light on the role and importance of magnetic correlations in these compounds. In particular, we find no evidence for strong magnetic correlations in  superconducting \LaOFFeAs.
 NMR and NQR are well suited to probe magnetic correlations,  since they are sensitive local probe techniques giving access to the intrinsic susceptibility, with a nucleus-dependent sensitivity to certain regions in $\vekg{q}$-space.
In Sec.~\ref{str:experimental}, we outline the experimental details of sample preparation and characterization, as well as NMR- and theory-related details.
The NMR results on the electronic spin susceptibility are then presented and discussed in Sec.~\ref{str:spinsuscept}.
Our Knight shift and spin lattice relaxation rate measurements of three different nuclei (\as , \fe, and \la ) scale with one another as a function of temperature. This result is surprising, as it suggests that all nuclei couple to the same spin degree of freedom, and that there is little or no $\vekg{q}$-space structure in the dynamical spin susceptibility.  If spin fluctuations were present, with a correlation length greater than about one lattice spacing, then the \as\ and \la\ spin lattice relaxation rates would differ from that of the \fe , in contrast with our observations. We also measure the Knight shift of the \as\ up to 480 K, in order to investigate the behavior of the spin susceptibility at high temperature.  We find that the shift increases monotonically up to 480 K, showing no sign of a peak as observed in pseudogap studies of the cuprates \cite{Curro1997}.
Finally, in Sec.~\ref{str:NQR}, we present density functional calculations of the spatial charge distribution and electric field gradient (EFG) at the \as\ site  in \LaOFFeAs\ and \LaOFeAs\ and compare with our experimental observations.

\section{Experimental details and theoretical methods}
\label{str:experimental}

\subsection{Sample preparation and characterization}

Polycrystalline samples of \LaOFxFeAs\ with $x$=0.1 and $x$=0 were prepared by standard methods and characterized by x-ray diffraction, resistivity and susceptibility measurements as described in \cite{Grafe2008,Kondrat2008}.
A value of $T_c \approx$ 26.0 K was extracted from these measurements for a fluorine doping of $x$=0.1.
A similar but \fe-enriched sample showed a reduced $T_c \approx$20 K from low-field SQUID measurement, with a reduced Meissner effect.
The origin of the reduced $T_c$ is not yet clear. A change in the doping level could be excluded, as the \as\ NQR spectrum was unchanged. The undoped (non-superconducting) sample exhibits a structural phase transition from tetragonal to orthorhombic at $T_S \approx 156$ K followed by a spin density wave transition at $T_N \approx 138$ K \cite{Luetkens2008}.
For the NMR experiments, an oriented powder of the (unenriched) $x=0.1$ sample was formed by grinding the material to a powder, mixing with Stycast 1266 epoxy, and curing in an external field of 9.2 T.

\subsection{NMR and NQR}

NMR is a powerful probe of the  behavior of the electronic system in the ferropnictides because the nuclei interact with the electrons via quadrupolar and hyperfine interactions. The nuclear Hamiltonian is given by:
\begin{equation}
{\cal H} = -\gamma_n \hbar  I \cdot (1+K) H_0 +
\frac{h\nu_Q}{2}\left[I_{z}^2-\frac{I(I+1)}{3}+\frac{\eta}{6}(I_+^2+I_-^2)\right],
\label{hamiltonian}
\end{equation}
where $\gamma_n$ and $I$ are the gyromagnetic ratio and the spin of the nucleus, $\hbar$ is Planck's constant, $H_0$ is the applied magnetic field, $K$ is the NMR shift, and $\nu_Q$ and $\eta$ the quadrupole frequency and the asymmetry of the electric field gradient tensor.  The quadrupole moment, $Q$, of nuclei with spin $I > 1/2$ (La, As) interacts with the electric field gradient (EFG), which depends on the local charge symmetry. The quadrupolar interaction lifts the degeneracy of the $I$ multiplet, giving rise to  $2I+1$ resonances in an applied field (NMR), and $I-1/2$ resonances in zero field (NQR).

The Knight shift $K$ arises through the hyperfine interaction between the spins of the electrons and the spins of the nuclei. In the presence of an applied field (NMR), the static field of the polarized electrons creates an additional field at the nuclear sites, yielding a shift, $K$, of the resonance line with respect to the unshifted Larmor frequency. Furthermore, any time dependence of this hyperfine field (due to electronic spin and orbital moment fluctuations), gives rise to spin lattice relaxation. In general, $K$ can be written as the sum of a temperature-dependent spin part, $K_s$, and a usually temperature-independent orbital part, $K_{orb}$.
$K_s$ is proportional to the static susceptibility of the electrons at the Fermi level, $\chi_s(q=0)$:
\begin{equation}
K_s = A_{hf} \cdot \chi_s(q=0) \label{eqn:Ks}
\end{equation}
where $A_{hf}$ is the hyperfine coupling, which depends on the nucleus through the availability of various coupling paths, i.e., on-site coupling or transferred coupling to neighbouring atoms.

\begin{table}
\caption{\label{tab:nuclei} NMR/NQR properties of the studied nuclei. The relative sensitivity for iron assumes 100\% \fe\ enrichment.}
\begin{indented}
\item[]\begin{tabular}{@{}cccccc} \br
 & $I$ & $\gamma_n$ & $Q$ & natural abundance & relative sensitivity \\
 & & (MHz/T) & ($10^{-28}$~m$^2$) & (\%) & ($^1H$ = 1) \\
\mr
\la & 7/2 & 6.014\phantom{0} & 0.2\phantom{00} & $\approx$100 & 5.9$\cdot 10^{-2}$ \\
\fe & 1/2 & 1.3757 & 0\phantom{.000} & $\approx$2.2 & 3.4$\cdot 10^{-5}$ \\
\as & 3/2 & 7.2917 & 0.314 & \phantom{$\approx$}100 & 2.5$\cdot 10^{-2}$ \\
\br
\end{tabular}
\end{indented}
\end{table}

All of the nuclei are NMR-active in \LaOFFeAs; we have chosen to focus on \la, \as, and \fe, whose properties are given in Tab.~\ref{tab:nuclei}.
Since the La is located out of the FeAs plane, it is expected to be coupled only weakly  to the electronic spin system.
The EFG at the La site is particularly small, rendering NQR experiments difficult to carry out.
On the other hand, As is located directly in the FeAs planes, and therefore should have a  comparatively higher coupling to the electronic spins.
Moreover, the combination of a large quadrupole moment and EFG allows for direct NQR experiments on the \as. The spin 1/2 nature of the \fe\ is ideal for contrasting with the \as\ in order to distinguish quadrupolar from magnetic effects. However, the low natural abundance of the \fe\ isotope makes it necessary to enrich the sample, using \fe\ during the synthesis. Nevertheless, the sensitivity remains low compared to the other nuclei, making these experiments more difficult.

When doing NMR on a powder sample, the crystallites are randomly oriented with respect to the applied magnetic field. This implies the random orientation of the magnetic and quadrupolar electric hyperfine tensors, giving rise to broad spectra and reduced signal intensities.
Although EFG and magnetic shift tensors can be extracted from powder patterns in principle, we have chosen to exploit the anisotropic character of the magnetic susceptibility of these materials.
By letting a mixture of powder and liquid epoxy cure while subjected to an external field, one obtains a powder with its crystallites oriented along the direction of highest susceptibility.
In the case of \LaOFFeAs\ the susceptibility is larger in the $ab$ plane, i.e., along two directions.
Therefore, we obtain $ab$ oriented samples where the $ab$ planes are parallel to the applied field, with a distribution of the resonance simpler (2D powder) than in the standard powder case.

\subsection{LDA calculations}

The band structure calculations were performed using the full-potential local-orbital minimum basis code FPLO (version 5.00-19) \cite{FPLO} within the local density approximation (LDA).
In the scalar relativistic calculations the exchange and correlation potential of Perdew and Wang \cite{PW} was employed.
As basis set La (5s5p/6s6p5d+4f7s7p), Fe (3s3p/4s4p3d+4d5s5p), As (4s4p3d+4d5s5p) and O (2s2p3d+3s3p) were chosen for semicore/valence+polarization states.
The high lying states improve the basis which is especially important for the calculation of the electric field gradient (EFG) tensor with the components $V_{ij}=\partial V/\partial x_i\partial x_j$.
The lower lying states were treated fully relativistic as core states.
A well converged $k$-mesh of 275 $k$-points was used in the irreducible part of the Brillouin zone. LaFeAsO was calculated in space group 129 (P4/nmm) with the structural parameters as given in \cite{Cruz2008}. In order to investigate the influence of F substitution on the O site, the virtual crystal approximation (VCA) was applied and cross checked with the calculation
of super cells \footnote{In order to come close to the experimental F
  concentration of 10~\%, a $4-$fold super cell (doubled along $a$ and
  $b$), with 8 formula units was calculated. Therefore the space group
  $P$mm2 (\#25) was chosen. Replacing one O by F yields a composition
  of LaFeAsO$_{0.875}$F$_{0.125}$.}.

\section{Spin susceptibility}
\label{str:spinsuscept}

In this section, we present results pertaining to the electronic spin susceptibility of the FeAs layers, as probed by \la, \as, and \fe\ nuclei.
Through spectra and relaxation measurements, we probe the temperature dependence of the static and dynamic susceptibility.

\subsection{Uniform susceptibility}

\subsubsection{Knight Shift Tensor at the  \fe}

\begin{figure}
\begin{center}
 \includegraphics[width=\columnwidth,clip]{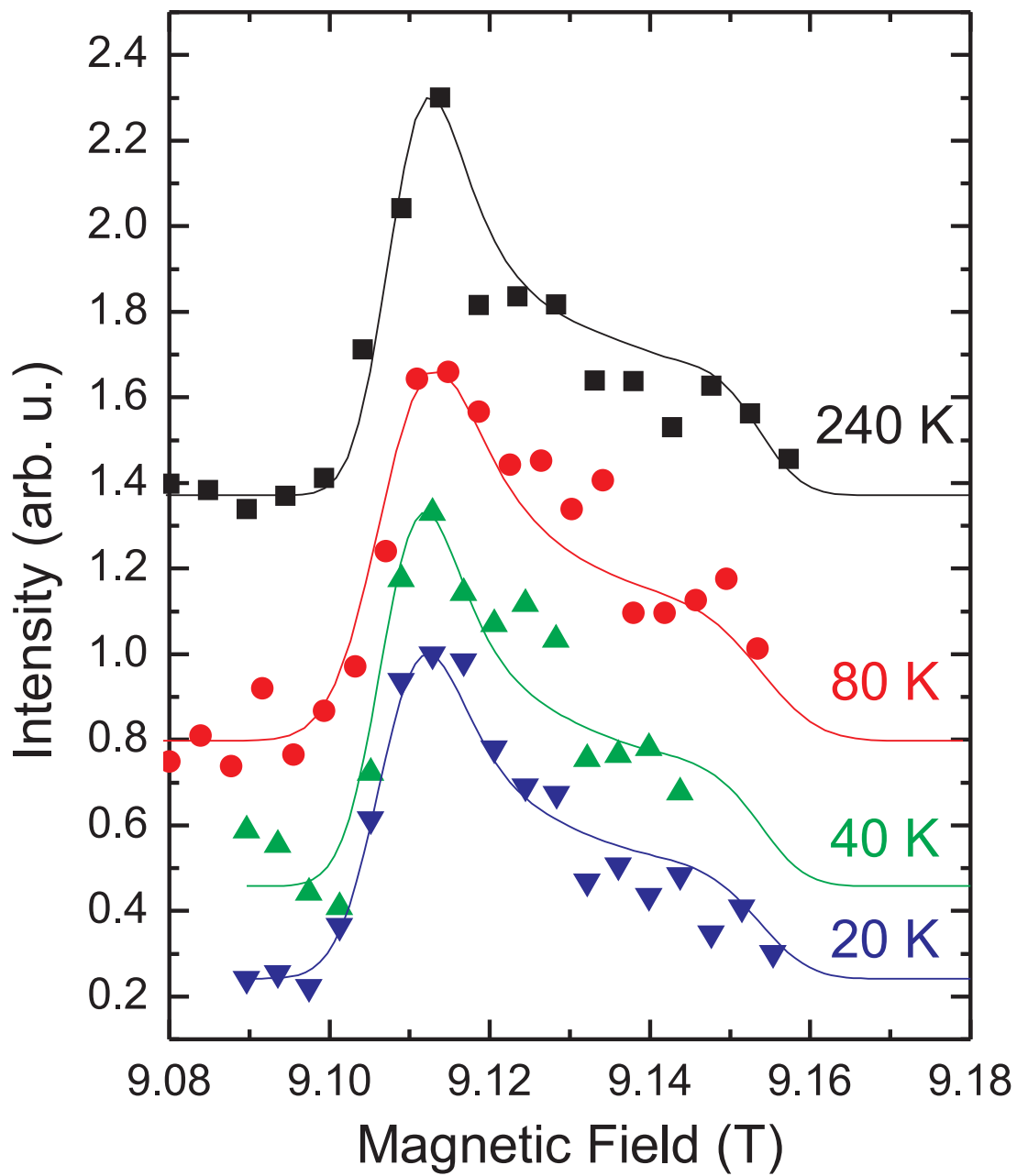}
 \caption{Field sweeped NMR spectra of the \fe\ for different temperatures measured at an applied frequency of 12.7 MHz. The full lines are simulations as described in the text.}
 \label{fespec}
\end{center}
\end{figure}

Accessing the temperature dependence of the susceptibility at $\vekg{q}=0$ was done through measurements of the NMR shift for the three studied nuclei, as derived from fitting of experimental spectra.
A representative example is given on Fig. \ref{fespec}, which shows \fe\ NMR data for an unaligned doped LaO$_{0.9}$F$_{0.1}$FeAs powder.
The data was obtained using the standard spin echo sequence with $\tau$=24 $\mu$s, while sweeping the field at a constant frequency of 12.7 MHz.
These spectra correspond to the iron (1/2 $\leftrightarrow$ -1/2) transition at several temperatures.
For $T$=20 K ($T_c < 20$ K at $H_0=9.2$ T, see Sec.~\ref{str:experimental}), we can clearly resolve a high field tail in the spectrum. Since \fe\ has $I=1/2$, there are no quadrupolar contributions to this spectrum, and in fact this shape is typical of a powder distribution with an anisotropic Knight shift tensor. This interpretation is confirmed by a simulation of such a distribution (see Refs \cite{Narita1966,Creel1974}), shown as a solid line in Fig. \ref{fespec}, allowing  us to extract the eigenvalues $K_{1/2/3}$ of the magnetic hyperfine tensor.
We find $K_1=K_2=1.35(1)$ \% and $K_3=0.85(2)$ \%, with the higher uncertainty on $K_3$ due to the lack of definition of the end of the high-field tail ($K_{1/2}$ relate to the low-field peak).
In light of the layered structure, it is reasonable to assume that the principal axes
of the magnetic hyperfine tensor lie along the $c$ axis and within the $ab$ plane, with equal in-plane eigenvalues, and we assign $K_{1/2}$ and $K_3$ as $K_{ab}$ and $K_c$.
On increasing the temperature, we observe similar spectra, at least for the low-field peak (the poor sensitivity precludes comparison at higher field).
The full lines show again powder simulations, with $K_c$ held at its value at $T=20$ K so that the high-field decrease remains the same.
The low-field peak, which is solely determined by $K_{ab}$, shows a weak temperature dependence (see Fig. \ref{Kandchi}).
If we assign the temperature-independent component to the orbital shift (as is customarily done), we find a quite small (less than 0.1\%) spin component of the NMR shift along $ab$, with most of the shift being of orbital origin ($K^{ab}_{\rm orb} \approx 1.4$\%).
Note however the much smaller value of $K_c$ compared to $K_{ab}$.
In a similar \fe\ study on LaFeAsO$_{0.7}$, Terasaki \etal\ \cite{Terasaki2008} measure an even bigger anisotropy at $T$=30~K, with a shift of roughly 0.5\% along $c$ compared to again 1.4\% in the plane.
One possibility would be that this strong difference between the $c$ and $ab$ directions originates from a large difference in spin shift, well exceeding an order of magnitude, with an isotropic orbital shift.
This would imply a very large anisotropy of the iron hyperfine coupling, since SQUID measurements rule out any such large anisotropy of the susceptibility. Having an isotropic orbital shift in spite of the doping would also likely imply at minimum significant mixing of the $3d$ orbitals.
While we cannot rule out that scenario, a more straightforward explanation would be that there is a large orbital shift anisotropy (implying $K^{c}_{\rm orb} \approx 0.8$\%), reflecting the fact that the iron valence shell is not closed.
This may yield indications on the average valence of the iron, thus on the doping, as was suggested by Mukhamedshin \etal\ \cite{Mukhamedshin2005} in the study of sodium cobaltates.

\subsubsection{Scaling and Evidence for a Single Component}

\begin{figure}
\begin{center}
 \includegraphics[width=\columnwidth,clip]{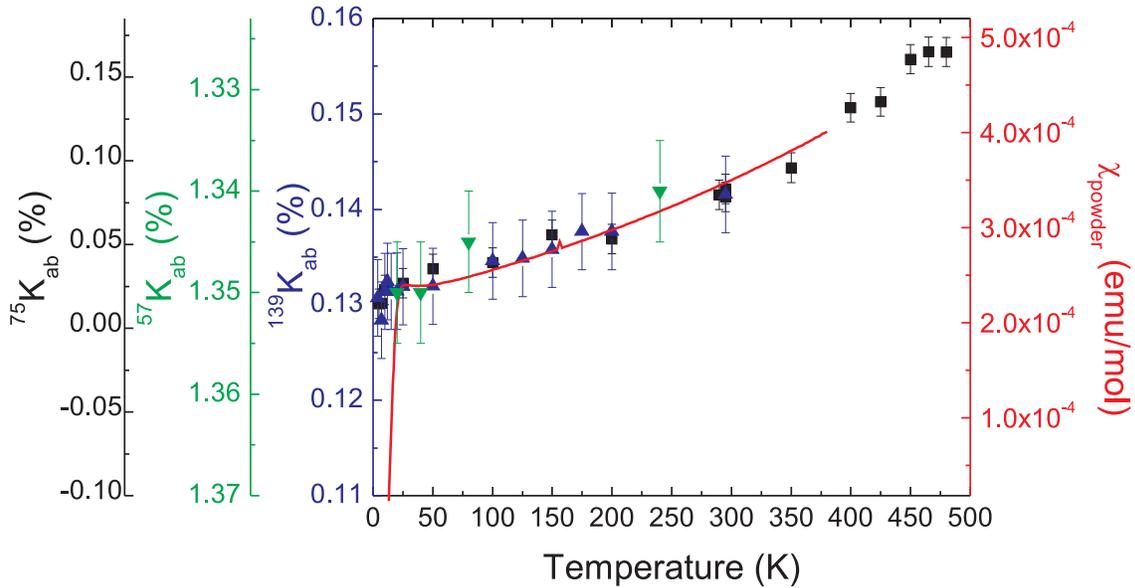}
 \caption{Knight shift of \as\ (black squares), \fe\ (green down-pointing triangles) and \la\ (blue triangles), and the macroscopic susceptibility $\chi_{powder}$ (solid red line)
 versus temperature with different vertical scales and origins. Note the reversed scale for \fe.}
 \label{Kandchi}
\end{center}
\end{figure}

The extracted temperature dependence of the \fe\ shift $K_{ab}$ is presented on Fig. \ref{Kandchi}.
Also shown are the \as\ (extended to high temperatures from previous work \cite{Grafe2008}) and \la\ Knight shift in the $ab$ directions, \Kab, as measured in oriented \LaOFFeAs\ powder (2D powder) with a field $H_0$=7.0494~T applied along the $a,b$ directions.
The resonance frequencies of the central transitions of the As and La are given to second order by $f = \gamma H_0 (1+K_{ab}) + 3\nu_Q^2/16\gamma H_0$, with $f$ the frequency of the peak of the 2D powder spectrum.  For \as, we have independently measured $\nu_Q = 11.00(5)$ MHz (see Fig. \ref{nuQ} and Ref. \cite{Grafe2008}), and therefore we can extract $K_{ab}$ from the spectrum. For the La, we measured the position of the satellites in a full spectrum, and found that  $\nu_Q = 1.15(5)$ MHz.
The macroscopic powder susceptibility measured in an applied field of 5 T is shown as a solid line.
By proper scaling, the four data sets can be made to overlap in the paramagnetic region.
The legitimacy of such a procedure is based on the fact that each shift can be written as the sum of a temperature-independent orbital shift $K_{orb}$, plus a spin shift $K_s$. Likewise, the static susceptibility can be written as the sum of a diamagnetic term $\chi_{dia}$ plus a Van Vleck-like term $\chi_{VV}$, both expected to be temperature-independent, and a temperature-dependent spin term $\chi_{s,macro}$.
Therefore the ratio of the scales of the NMR shifts to the scale of $\chi_{powder}$ reflects the strength of the hyperfine coupling of each nucleus to the spin susceptibility. It is remarkable that such general scaling can be obtained, and suggests that the three nuclei probe the same component of spin susceptibility. 

\textit{A priori} this result is surprising, as there are several bands crossing the Fermi surface in the ferropnictides, and one might expect each band to supply a different contribution to the spin susceptibility with different hyperfine couplings to different bands. Such is the case for the oxygen in Sr$_2$RuO$_4$, which simultaneously couples to multiple bands with different temperature dependent susceptibilities \cite{ImaiSr2RuO4}. The Mila-Rice-Shastry picture in the cuprates, however, captures much of the relevant physics in terms of a single spin component \cite{Takigawa1991,Alloul1993,MilaRice1989,Shastry1994}. Our results suggest that if there are different hyperfine couplings to multiple orbitals in the ferropnictides, the spin response of each of it is nearly identical. This observations may reflect the itinerant, rather than localized character of the Fe 3d electrons.

The shift data in Fig. 2 clearly show a strong decrease of the spin susceptibility with decreasing temperature, in agreement with several previous reports \cite{Ahilan2008,Grafe2008,Klingeler2008}.
This suppression of low-energy spin excitations is similar to the behavior of the cuprates, and hence has been ascribed to the existence of a pseudogap in this system.
In the cuprates, the spin susceptibility reaches a maximum at a temperature $T^*$ that is doping dependent. Since our Knight shift measurements indicate an increasing susceptibility with increasing temperature, we sought to find if this trend continues to higher temperature. Our \as\  measurements up to 480 K show no signature of a pseudogap peak, although the data hint at a leveling out by 500 K. It is unclear whether points at even higher temperatures could be gathered, as was for instance fruitful for YBa$_2$Cu$_4$O$_8$ \cite{Curro1997}, since SQUID measurements suggest a degradation of the sample.
Note that the scaling of the NMR shift with $\chi_{powder}$ remains good down to $T_c$, which indicates the high quality of the samples, and the absence of any signature of a Curie contribution due to paramagnetic impurity spins, either intrinsic or belonging to a spurious phase.

\subsubsection{Hyperfine couplings}

Table \ref{tab:AhfKorb} summarizes the hyperfine couplings and temperature-independent shifts extracted by plotting the measured shifts versus the bulk susceptibility, using Eq. \ref{eqn:Ks}. Since the anisotropic components of $\chi$ are unavailable in our aligned powder, we compare with the powder susceptibility in this determination, ignoring the modest anisotropy in temperature dependence between the $c$ and $ab$ directions \cite{chen2008}.
We do not have an estimate for the non-spin component $\chi_{VV} + \chi_{dia}$ but it is between an overly-cautious lower bound of 0 and an upper bound of roughly $2 \cdot 10^{-4}$ emu/mol, otherwise the spin contribution would have to become negative below a certain temperature.
This gives for the temperature-independent fraction of the shift $^{139}K_{orb}^{ab} = 0.12(1)$\% and $^{57}K_{orb}^{ab} = 1.36(1)$\%.

\begin{table}
\caption{\label{tab:AhfKorb}Hyperfine couplings and orbital shifts for \la, \fe, and \as\ in the \LaOFFeAs\ compound. The results for \as\ are from previous work \cite{Grafe2008}.}
\begin{indented}
\item[]\begin{tabular}{@{}lccc} \br
 & \la & \fe & \as \\
\mr
$A_{hf}^{ab}$ (kOe/$\mu_B$) & 4.3(8) & -5.7(14) & 25(3) \\
$K_{orb}^{ab}$ (\%) & 0.12(1) & 1.36(1) & -0.03(4) \\
\br
\end{tabular}
\end{indented}
\end{table}

The largest hyperfine coupling is to the As, which may be due in part to the fact that there are four nearest neighbor iron atoms to each As.  The fact that $^{139}A_{hf}$ is roughly six times lower than ${^{75}A_{hf}}$ is not surprising given that lanthanum is outside of the FeAs layers.
In the case of iron, the negative hyperfine coupling suggests that the dominant hyperfine coupling is via a core polarization mechanism, but the small magnitude of the Fe hyperfine coupling is surprising.  \textit{A priori}, one would expect that iron would have the strongest coupling to the electronic properties, in light of theoretical predictions \cite{Singh2008,Ma2008,Vildosola2008} indicating the highly predominant iron $3d$ character of the bands at the Fermi level. One explanation is that $^{57}A_{hf} = A_{cp} + A_{4s}$ where $A_{cp}$ (the core polarization contribution) is large and negative, whereas $A_{4s}$ is large and positive, so the net coupling remains small.

From the measured lanthanum coupling, it is possible to estimate the value of the iron moment $m_{Fe}$ in the magnetically-frozen SDW state of the undoped LaOFeAs material. As there are no major structural changes between the 10\% fluorine-doped material and the undoped material in its low-temperature state (aside from the slight orthorhombic distortion as a prelude to  the magnetic transition) it is reasonable to assume that $^{139}A_{hf}$ remains doping independent.  In this case, using the internal field of $H_{int}(La) = 2.5$ kOe measured by Nakai \etal\ at the lanthanum site in the SDW state \cite{Nakai2008} and the relation $H_{int}(La) = {^{139}A_{hf}} \, m_{Fe}$, we estimate $m_{Fe} = 0.58(9) \mu_B$. While this is larger than the values obtained through M\"ossbauer \cite{klauss} and neutron scattering \cite{Cruz2008} measurements, which gave respectively $m_{Fe}=$0.25 $\mu_B$ and $m_{Fe}=$0.36 $\mu_B$, this result still points clearly at a largely itinerant situation in the SDW state.
%A possible source of discrepancy between the techniques would be an unexpected change in the value of $^{139}A_{hf}$, impacting NMR.

\begin{figure}
\begin{center}
 \includegraphics[width=\columnwidth,clip]{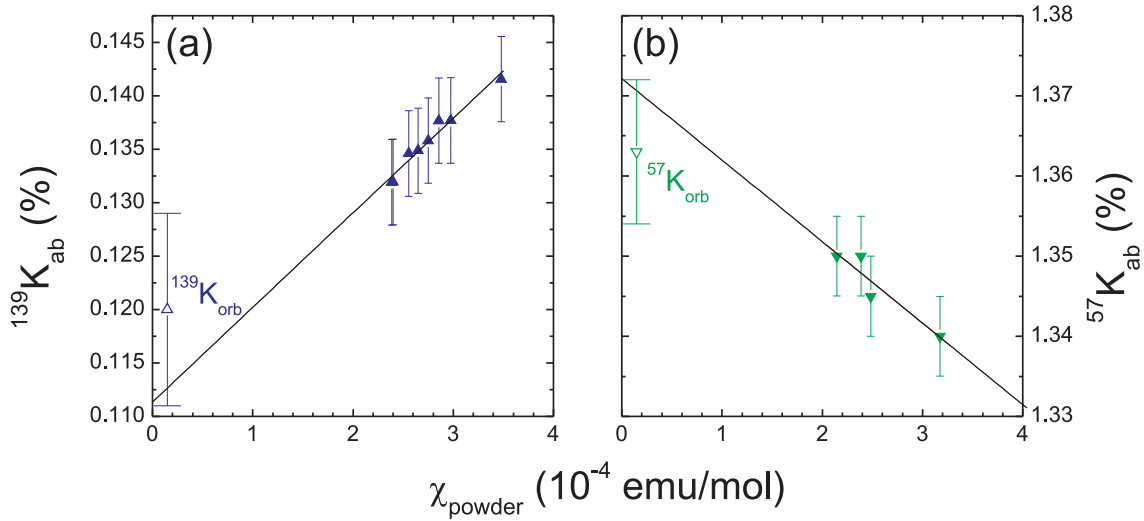}
 \caption{Knight shift of \la\ (a) and of \fe\ (b) versus the macroscopic susceptibility $\chi_{powder}$. The black lines are linear fits. The orbital shifts (see text) are shown next to the left vertical axes.}
 \label{Lavschi}
\end{center}
\end{figure}

\subsection{Wavevector dependence of the susceptibility}

As the superconductivity and frozen magnetism regions are close to each other in the phase diagram, it is natural to consider that spin fluctuations may play a role in the formation of Cooper pairs. NMR measurements of the spin lattice relaxation \slrr\ sample the low energy spin fluctuations via the dynamic susceptibility $\chi''$  \cite{Moriya1963}:
\begin{equation}
 T_1^{-1} \propto \frac{\gamma_n^2k_BT}{\gamma_e^2}\lim_{\omega\rightarrow0}\sum_{\vekg{q}} |A(\vekg{q})|^2\frac{\chi''_{\perp}(\vekg{q},\omega_0)}{ \omega_0 },
\label{T1}
\end{equation}
where $k_B$ is Boltzmann's constant, $\gamma_{e/n}$ the gyromagnetic ratios of the electron and the probed nucleus, $\omega_0$ the Larmor frequency, and $A(\vekg{q})$ the hyperfine form factor of the probed nucleus.
Particular attention should be paid to the $\vekg{q}$-dependence of the latter, as filtering effects may occur such as in the superconducting cuprates \cite{Mila1989}, wherein the oxygens in the CuO$_2$ layers are insensitive to antiferromagnetic fluctuations, whereas the copper nuclei are.
Here, the fact that \fe\ appears to be a particularly poor probe of the uniform ($\vekg{q}$=0) susceptibility compared to \as\ and \la, taking into account the crystallographic positions, suggests that hyperfine filtering effects might be at play.
While the numerous superexchange paths make the analysis difficult for fluorine and lanthanum, Terasaki \etal\ \cite{Terasaki2008} show indeed that $^{75}A_{hf}(\vekg{q})$ is well-developed at $\vekg{q}=0$ and tends to vanish towards $\vekg{q}=(\pi/a, \pi/a)$, while $^{57}A_{hf}(\vekg{q})$ exhibits an opposite behaviour, tending to low values around $\vekg{q}=0$.
%In practice, such significant differences between the nuclei can be exploited through measurement of the spin lattice relaxation rate, \slrr.

\begin{figure}
\begin{center}
 \includegraphics[width=\columnwidth,clip]{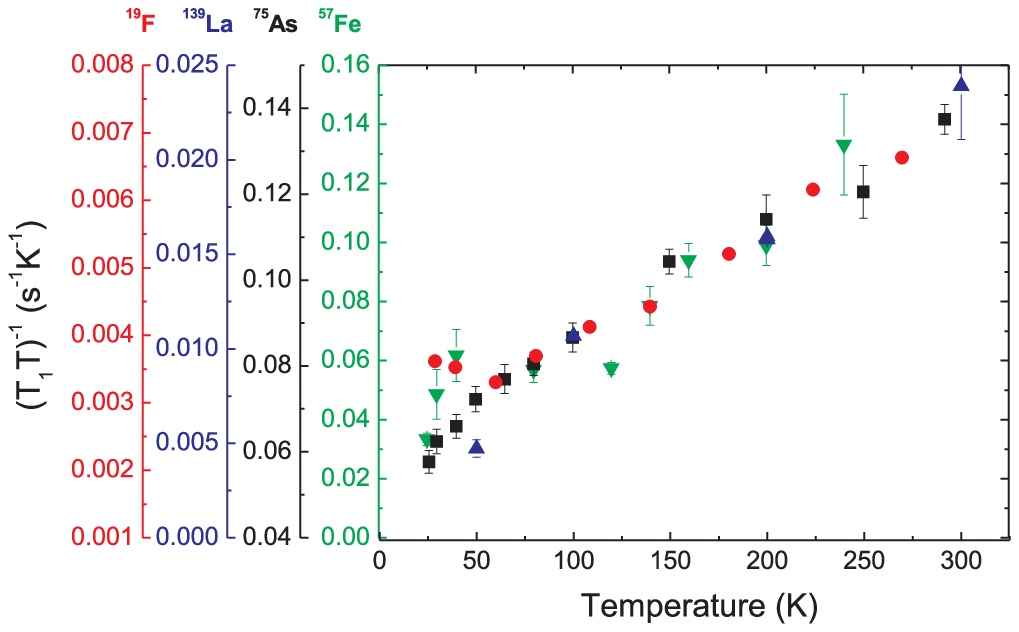}
 \caption{The temperature dependence of the \fe, \as, \la, and \f\ \slrrt. The \f\ data is reproduced from Ahilan \etal\ \cite{Ahilan2008}. For \la, \fe, and \as, $T_1$ along the $ab$ directions is used, while for fluorine it is $T_1^{iso}$.}
 \label{T1all}
\end{center}
\end{figure}

The temperature dependence of \slrrt\ for each of the Fe, As and La nuclei is shown on Fig. \ref{T1all}, including \f\ data from Ahilan \etal\ \cite{Ahilan2008} for comparison. For \la, \fe, and \as, $T_1$ along the $ab$ directions is used, while for fluorine it is $T_1^{iso}$.
The data are plotted with different axes, in order to highlight the similar temperature dependence for all four nuclei. The strength of the relaxation correlates with the distance to the iron plane, with the lowest values for the nuclei outside the FeAs planes. The \slrrtext\ is largest for the Fe, whereas the spin shift at the Fe is relatively small compared to the other sites.  In other words,  $^{75}(T_1 T \gamma_n^2) / ^{57}(T_1 T \gamma_n^2) \approx$20--30 while $(^{57}A_{hf})^2 (\vekg{q}=0) / (^{75}A_{hf})^2 (\vekg{q}=0) \approx$0.05. There are two possible explanations for this discrepancy.  Either there is a  strong $\vekg{q}$-space dependence to $\chi"(\vekg{q},\omega)$, or there are multiple hyperfine coupling channels ($A_{cp}$, $A_{4s}$) between the Fe nuclear moments and the same degree(s) of spin freedom.
Given the fact that the \slrrtext\ for all the nuclei exhibit roughly the same temperature dependence as seen for the NMR shifts, any strong $\vekg{q}$ dependence to the dynamic susceptibility seems unlikely.  In other words, while the susceptibility is temperature-dependent, the decrease occurs similarly across $\vekg{q}$-space. 
While some fluctuations in certain $\vekg{q}$ regions cannot be ruled out in the absence of a refined analysis of hyperfine filtering effects, a simple explanation of our data would be that the relaxation comes mostly from quasi-particle scattering. In this case, the core-polarization and the diamagnetic contributions to the Fe relaxation rate add as the sum of the squares rather than a direct sum, as discussed in \cite{Curro2003}. We note that there are no signatures of a pseudogap peak in the \slrrt\ data, as observed in the cuprates \cite{Takigawa1991,Berthier1996} in either the Fe or As data.

\section{Spatial charge distribution}
\label{str:NQR}

As a complementary study, we present in this section a theoretical analysis of the issue of the spatial charge distribution, based on our \as\ NQR measurements. The EFG observed at the As site in the ferropnictides varies dramatically from one compound to the next, in stark contrast to the cuprates, where the EFGs are slightly doping dependent, but exhibit little variation among different families. To address this, we have measured the doping dependence of the NQR spectrum, and compared with LDA calculations.

\subsection{NQR results}

\begin{figure}
\begin{center}
 \includegraphics[width=\columnwidth,clip]{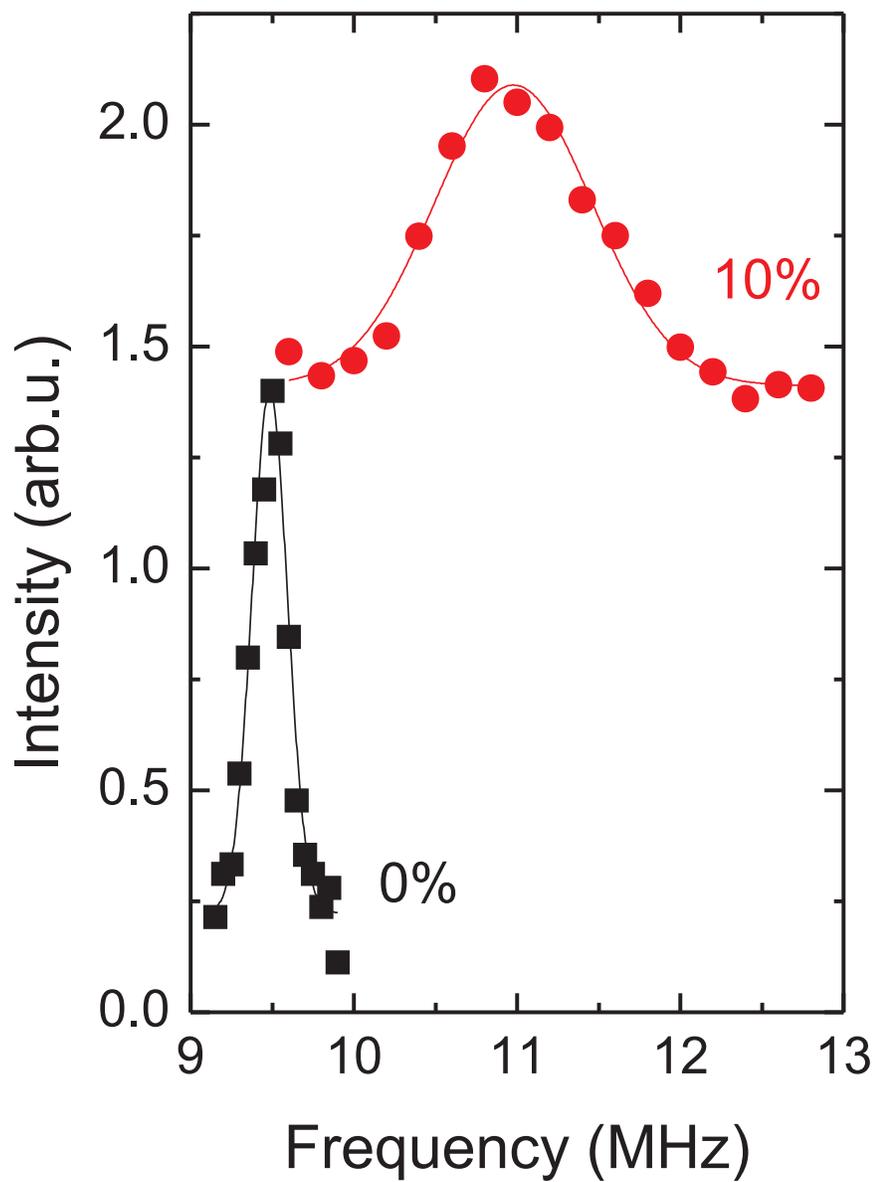}
 \caption{Room temperature \as\ NQR spectra in the doped \LaOFFeAs\ material and in the undoped parent compound LaOFeAs. Solid lines indicate gaussian fits, with parameters given in the text.}
 \label{fig:NQRspec}
\end{center}
\end{figure}

In the ferropnictides, the \as\ nucleus quadrupolar moments interact with the electric field gradient (EFG) tensor giving rise to a resonance at frequency:
\begin{equation}
\nu_Q = \frac{3eQV_{zz}}{2I(2I-1)h} \sqrt{1+\eta^2/3} \label{nuQ}
\end{equation}
where $\eta$ is the asymmetry of the EFG tensor.
The EFG has the symmetry of the local atomic position, and depends on the local electronic density.  
We present on Fig. \ref{fig:NQRspec} the \as\ NQR spectrum at room temperature for the doped \LaOFFeAs\ material studied here above, as well as for the undoped parent compound LaOFeAs.
%As the \as\ nuclear spin is 3/2, one expects one resonance line for every unique charge environment at the As sites.
For both dopings, a well defined line is observed, meaning that in each sample the EFG is the same at all As sites.
This is in agreement with the single As crystallographic site and indicates spatially homogeneous doping.
In the doped sample, the line is however significantly broader (full width at half maximum of 0.97(6) MHz) than in the undoped case (FWHM=0.22(1) kHz).
This could be explained by limited inhomogeneities of the fluorine concentration in the material, distributing somewhat the EFG, or even simply by the fact that As ions are at varying distances from the fluorine.
%For a nuclear spin 3/2, the presence of only one line makes it impossible to extract $\eta$.
%We know however from the \LaOFFeAs\ NMR spectrum that $\eta \approx 0.1$.
%The even more pronounced axial symmetry at the \as\ site in the undoped case, due to the absence of any fluor ion, suggests that $\eta$ is likely similar or even smaller in that case.
We find that $\nu_Q$ increases from 9.48(1) MHz to 11.00(5) MHz on doping, which 
translates accordingly in a 16\% increase of $V_{zz}$.  This trend is in good agreement with measurements by Mukuda \etal\ \cite{Mukuda2008} on oxygen-deficient compounds.

\subsection{Comparison with theory}

\begin{figure}
\begin{center}
 \includegraphics[width=\columnwidth,clip]{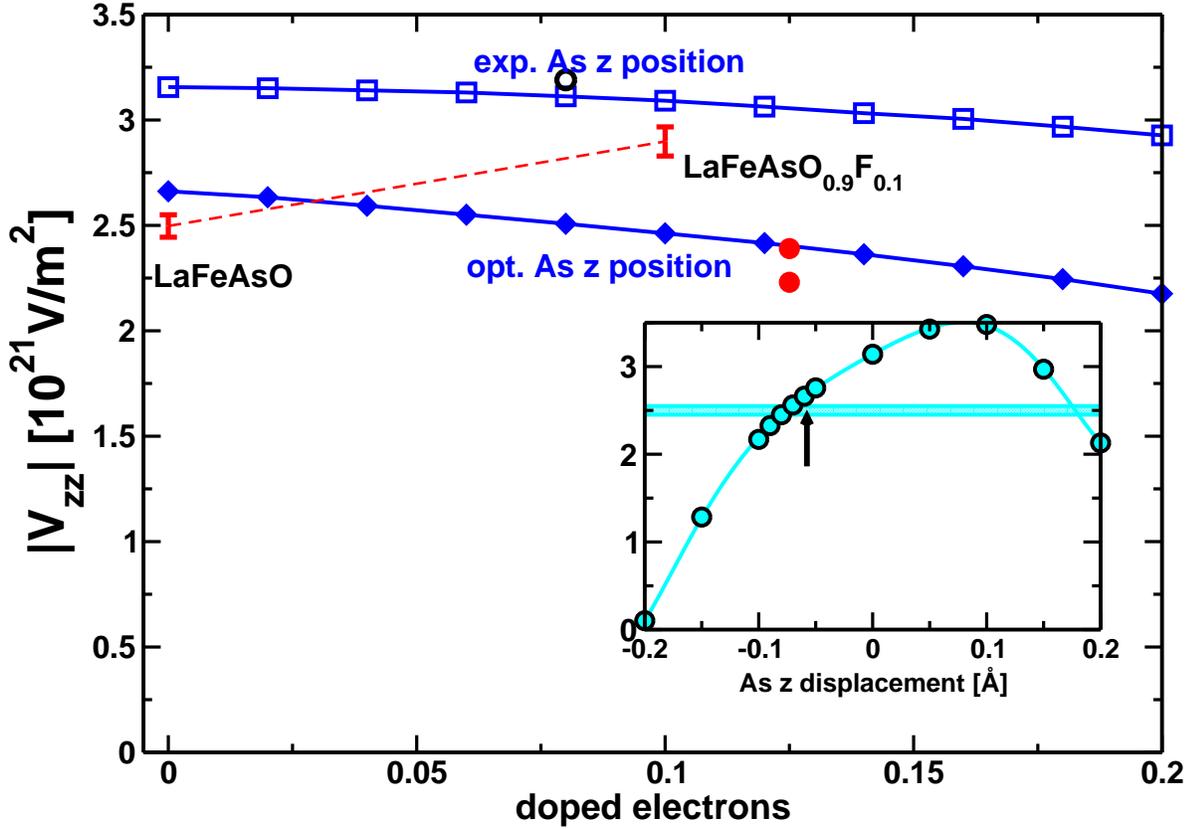}
 \caption{Calculated $V_{zz}$ obtained from the virtual crystal approximation using the experimental As $z=0.6507$ position (empty squares), the optimized As $z=0.6438$ position (filled diamonds) and the LaFeAsO$_{0.92}$F$_{0.08}$ structure at 175~K (black circle). The red filled circles show the EFGs from the super cell calculation for the optimized As $z$ position. The measured EFGs for the pure and the 10~\% F-doped compound are shown by error bars. Inset: Dependence of $^{75}$As $V_{zz}$ on the As $z$ position. The energetically optimized As $z$ position is marked by an arrow. The experimental $V_{zz}$ is represented by the shaded bar.}
 \label{FigureEFG}
\end{center}
\end{figure}

%undoped
For the undoped compound the measured EFG is obtained by inserting the $^{75}$As quadrupole moment \cite{pyykko} $Q=(3.14\pm0.06)$~b in Eq.~(\ref{nuQ}), using $\eta$=0, which yields $|V_{zz}^{exp}| = (2.50\pm0.05) \cdot 10^{21}$~V/m$^2$.
Using the 175~K lattice parameters and atomic positions as given in Ref.~\cite{Cruz2008} we obtain a fair agreement for the calculated EFG: $V_{zz}^{calc}=-3.14 \cdot 10^{21}$~V/m$^2$.
Like for the Fe magnetic moment, which in the calculations shows a strong dependence on the As $z$ position \cite{CK, KK}, we observe also for the EFG a strong As $z$ dependence (inset to Fig.~\ref{FigureEFG}).
The calculated EFG ($V_{zz}^{calc} = -2.67\cdot 10^{21}$~V/m$^2$) is much closer to the measured value when As is shifted along the negative $z$-direction to $z=0.6438$, where the energy has a minimum (in inset to Fig.~\ref{FigureEFG} marked by an arrow) and the structure has a shorter Fe-As distance of $2.3748$~{\AA}.

%doped
The EFGs of the doped compounds were calculated with the virtual crystal approximation.
The validity of the VCA was confirmed by super cell calculations.
Due to the super cell construction, there are two different Wyckoff positions for As and hence two different EFGs, whereof one is lying on top of the VCA curve and the other one very close to it, see full red circles in Fig.~\ref{FigureEFG}.
First, we consider solely the effect of electron doping.
Therefore, we keep the structural parameters fixed for different levels of doping.
In Fig.~\ref{FigureEFG} two such VCA curves are shown.
When the experimentally determined As $z = 0.6507$ position is used, the calculated and measured EFG values for 10~\% doping agree quite well.
Also S. Leb\`egue \etal\ found good agreement for the 10~\% doped compound using the WIEN2k code \cite{LaCalc}.
The VCA curve with the optimized As $z=0.6515$ position is shifted in the direction of smaller $|V_{zz}|$.
Now, we investigate the structural change on top of the doping by calculating the EFG within VCA for the 175~K data of LaFeAsO$_{0.92}$F$_{0.08}$ as given in Ref.~\cite{Cruz2008}.
This has only a minor effect on the EFG, as it can be seen by the black circle in Fig.~\ref{FigureEFG}, which lies very close to the experimental VCA curve.
We stress at this point, that the effect of electron doping on the EFG is much smaller than the influence of the As $z$ position as can be clearly seen by comparing Fig.~\ref{FigureEFG} with the inset of Fig.~\ref{FigureEFG}.

Our calculations result in a decrease in $|V_{zz}|$ upon electron doping for LaFeAsO$_{0.9}$F$_{0.1}$, although in our experiments an increase is observed.
This is not pointed out by S. Leb\`egue \etal\ \cite{LaCalc}, although they obtain the same discrepancy
for the trend in the $V_{zz}$ calculation.
Further studies are required to investigate the connection with intrinsic changes in the electronic structure.

For the NdFeAsO system, note that there is a better agreement between the experimentally determined and the calculated EFG for the doped and for the undoped structure \cite{NdFeAsO}.
In that study, some of us (K. K., H. R) also showed, that the $4f$ electrons have only a minor influence on the EFG, whereas the structure (chemical pressure) influences the EFG to a higher degree.

\section{Conclusion}

Using simultaneous \la, \fe, and \as\ NMR and NQR measurements, we have investigated the electronic properties of the \LaOFxFeAs\ compound with x=0 and x=0.1. \as\ NQR measurements show a sizable evolution of the electric field gradient with doping that cannot be explained by LDA calculations, although the measured and calculated EFG are in reasonable agreement for the undoped parent compound. While a high sensitivity to the As $z$ position is observed, the electronic origin of the difference to the experimental spectra remains unclear. The systematic scaling of the NMR shifts with the macroscopic susceptibility in the paramagnetic state suggests the presence of a single spin degree of freedom. Extending \as\ shift measurements up to 480~K reveals a continuous increase of the spin susceptibility with no sign of a peak, suggesting that the observed pseudogap behavior persists up to at least this high. 
 We find that the hyperfine coupling to the iron is unexpectedly small, but speculate that there are in fact two components to the Fe hyperfine coupling with different signs.  Consequently the Knight shift at the Fe is rather small, but the spin lattice relaxation rate is large.  Furthermore, we find that the \slrrtext s for all four nuclei appear to scale with one another, providing further support for a single spin degree of freedom, and suggesting the absence of any significant structure in the dynamic susceptibility in $\vekg{q}$-space. This result is surprising: in contrast with the cuprates, there appear to be very little spin fluctuations at low energies present in these superconducting samples.  Detailed studies of the NMR relaxation and its doping dependence, as well as quantitative comparisons to inelastic neutron scattering data, should help bring further insight on the issue of the role of spin fluctuations in these materials. The origin of the strong temperature dependence of the susceptibility, as well as the presence of unconventional superconductivity in the absence of significant spin fluctuations remain open questions, and indicate that physics of the iron pnictides are very different from that of the cuprates.

\ack

We thank M. Deutschmann, S. M\"uller-Litvanyi, R. M\"uller, R. Vogel, and A. K\"ohler for experimental support. This work has been supported by the DFG, through FOR 538. D. P. acknowledges support from the DFG. G. L. acknowledges support from the Alexander von Humboldt-Stiftung.

\section*{References}
\bibliographystyle{unsrt}
\bibliography{Grafe}

\end{document}